\documentclass[11pt]{article}
\usepackage{amsmath}
\usepackage{amsfonts}
\usepackage{amssymb}
\usepackage{epsfig}
\usepackage{times}
\newcommand{\be}{\begin{equation}}
\newcommand{\ee}{\end{equation}}
\newcommand{\ba}{\begin{array}}
\newcommand{\ea}{\end{array}}
\newcommand{\bea}{\begin{eqnarray}}
\newcommand{\eea}{\end{eqnarray}}

\newcommand{\rar}{\rightarrow}

\renewcommand{\l}{\newline\null}
\def\figskip{\vskip .5cm plus 3mm minus 2mm}
\def\hbar{\not{\hbox{\kern-2.3pt $h$}}}
\def\psl{\not{\hbox{\kern-2.3pt $p$}}}
\def\Psl{\not{\hbox{\kern-2.3pt $P$}}}
\def\ksl{\not{\hbox{\kern-2.3pt $k$}}}
\def\qsl{\not{\hbox{\kern-2.3pt $q$}}}
\begin{document}
\begin{titlepage}
February 1999 (revised June 1999)\hfill PAR-LPTHE 99/06
\vskip 4.5cm
{\baselineskip 17pt
\begin{center}
{\bf THE DECAYS $\boldsymbol{\nu _H \rightarrow \nu_L\ \gamma}$
AND $\boldsymbol{\nu_H \rightarrow \nu_L\ e^+\ e^-}$ OF MASSIVE NEUTRINOS}
\end{center}
}
\vskip .5cm
\centerline{Q. Ho-Kim
     \footnote[1]{E-mail: qhokim@phy.ulaval.ca}
}
\vskip 2mm
\centerline{\em Department of Physics, Universit\'e Laval (Qu\'ebec)
     \footnote[2]{Sciences and Engineering Building, SAINTE FOY,
QC G1K7P4 (Canada).}}
\vskip 3mm
\centerline{B. Machet
     \footnote[3]{Member of `Centre National de la Recherche Scientifique'.}
     $^{,}$
     \footnote[4]{E-mail: machet@lpthe.jussieu.fr}
           \& X.Y. Pham
     \footnotemark[3]
     $^{,}$
     \footnote[5]{E-mail: pham@lpthe.jussieu.fr}}
\vskip 2mm
\centerline{{\em Laboratoire de Physique Th\'eorique et Hautes Energies,}
     \footnote[6]{LPTHE tour 16\,/\,1$^{er}\!$ \'etage,
          Universit\'e P. et M. Curie, BP 126, 4 place Jussieu,
          F-75252 PARIS CEDEX 05 (France).}
}
\centerline{\em Universit\'es Pierre et Marie Curie (Paris 6) et Denis
Diderot (Paris 7);} \centerline{\em Unit\'e associ\'ee au CNRS UMR 7589.}
\vskip 1.5cm
{\bf Abstract:}  If, as recently reported by the Super-Kamiokande
collaboration, 
the neutrinos are massive, the heaviest one $\nu_H$ would not be stable and,
though chargeless, could in particular decay 
into a lighter neutrino $\nu_L$ and a photon  by quantum loop effects.
The corresponding rate is computed in the standard model with massive Dirac
neutrinos as a function of the neutrino masses and mixing angles.
The lifetime of the decaying neutrino is estimated
to be $\approx 10^{44}$ years for a mass \hbox{$\approx 5\times 10^{-2}$ eV.}

If kinematically possible, the $\nu_H \rightarrow \nu_L \ e^+ \ e^-$ mode
occurs at tree level and its one-loop radiative corrections
get enhanced  by a large logarithm of the electron mass acting as
an infrared cutoff. Thus the $\nu_H \rightarrow \nu_L \ e^+  \ e^-$ decay
largely dominates the $\nu_H \rightarrow \nu_L \ \gamma$ one
by several orders of magnitude, corresponding to a lifetime $\approx 10^{-2}$
year for a mass $\approx 1.1$\;MeV.

\smallskip

\vfill
\null\hfil\epsffile{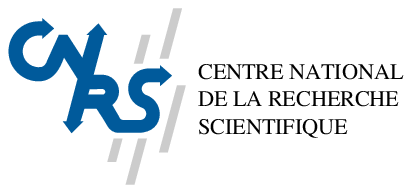}
\end{titlepage}
%
%
\section{Introduction}
\label{section:introduction}

Evidence for the transmutation between the two neutrino species $\nu_\mu
\leftrightarrow \nu_\tau$ has been recently reported  by the Super-Kamiokande
collaboration\cite{Superkamio}.
As a consequence, neutrinos could have non-degenerate tiny masses, and mixing
among different lepton families becomes likely, in analogy with the
Cabibbo-Kobayashi-Maskawa  flavor mixing in the quark sector \cite{CKM}.

We assume that the neutrino ``flavor'' eigenstates $\nu_e$, $\nu_\mu$
and $\nu_\tau$ are linear combinations of the three neutrino mass eigenstates
$\nu_1$, $\nu_2$ and $\nu_3$ of nonzero and non-degenerate masses
$m_1$, $m_2$ and $m_3$ respectively according to
\begin{equation}
\begin{pmatrix}\nu_e\cr \nu_\mu\cr \nu_\tau\cr\end{pmatrix}
=\begin{pmatrix} U_{e1} & U_{e2} & U_{e3}\cr
                 U_{\mu 1} & U_{\mu 2} & U_{\mu 3}\cr
                U_{\tau 1} & U_{\tau 2} & U_{\tau 3}\cr\end{pmatrix}
               \begin{pmatrix} \nu_1\cr \nu_2 \cr \nu_3\cr\end{pmatrix}
\equiv \;
{\cal U}_{lep}\;\begin{pmatrix} \nu_1\cr \nu_2 \cr \nu_3\cr\end{pmatrix},
\label{eq:Ulep}
\end{equation}
where the $3\times 3$ matrix ${\cal U}_{lep}$ is unitary.

The effective weak interactions of the leptons can now be written as
\begin{equation}
{\cal L}_{eff} =\frac{G_F}{\sqrt{2}} L^{\dagger}_\lambda L^\lambda,
\label{eq:Leff}
\end{equation}
where the charged current $L_\lambda$ is
\begin{equation}
L_\lambda =\sum _\ell\sum_{i =1}^{3} 
U_{\ell i}\overline{\nu}_i\gamma_{\lambda}(1-\gamma_5) \ell.
\label{eq:current}
\end{equation}
Here $\ell$ stands for $e^-$, $\mu^-$, $\tau^-$ and $\nu_i$ (with $i=1,2,3$)
are the three neutrino mass eigenstates.

Although the neutrinos are chargeless, a heavy neutrino $\nu_H$ can
decay into a lighter neutrino $\nu_L$ by emitting a photon; this
decay is entirely due to quantum loop effects. Now, if kinematically possible,
the mode $\nu_H \rightarrow \nu_L \ e^+ \ e^-$ largely dominates,
because it is governed by a tree diagram and its radiative corrections get
enhanced, as we will see, by a large logarithm.

Neutrino oscillation measurements provide constraints usually plotted in the
$(\sin 2\theta_{i j},\ \Delta m_{i j}^2 = |m_i^2-m_j^2|)$ plane, where
 $\theta_{i j}$ is one of the three Euler angles of the rotation matrix
${\cal U}_{lep}$.

For practical purposes, we shall assume for ${\cal U}_{lep}$ the following
form \cite{Peccei}:
\begin{equation}
{\cal U}_{lep} =
\begin{pmatrix}\cos\theta_{12} & -\sin\theta_{12}& 0 \cr
\frac{1}{\sqrt{2}}\sin\theta_{12} & \frac{1}{\sqrt{2}}\cos\theta_{12}  &
             -\frac{1}{\sqrt{2}}\cr
\frac{1}{\sqrt{2}}\sin\theta_{12} & \frac{1}{\sqrt{2}}\cos\theta_{12}  &
          \frac{1}{\sqrt{2}}\cr\end{pmatrix};
\label{eq:Upractical}
\end{equation}
$\theta_{23}\approx 45^0$ is suggested by the Super-Kamiokande data and
$\theta_{13}\approx 0^0$ comes from the CH00Z data \cite{Superkamio,Peccei}
which give $\theta_{13} \leq 13^0$, and also from the Bugey experiment
\cite{Bugey}, whereas $\theta_{12}$ is arbitrary.
Although  $\theta_{12}$ is likely small $\approx 0^0$, the maximal mixing
$\theta_{12}\approx 45^0$ may also be possible allowing
$\nu_e\leftrightarrow \nu_\mu$ (as suggested by the LSND experiment
\cite{Superkamio,Peccei}).

\section{The decay $\boldsymbol{\nu_H \rar \nu_L\ \gamma}$}
\label{section:photon}

The first calculations of radiative neutrino decays have been reported in
\cite{Petcov} and \cite{PalWolfenstein}.

In the most general renormalizable gauge (conventionally called $R_\xi$), six
Feynman diagrams contribute to the process  $\nu_H(P)\rar\nu_L(p)\ \gamma(q)$,
where the photon can be real ($q^2=0$) or virtual ($q^2\neq 0$); the latter
is necessary when we consider the one-loop radiative corrections to
$\nu_H\rar\nu_L\ e^+ \ e^-$. They can be grouped into two
sets: four in Figs.~1a-d:

\vbox{
\figskip
\epsfig{file=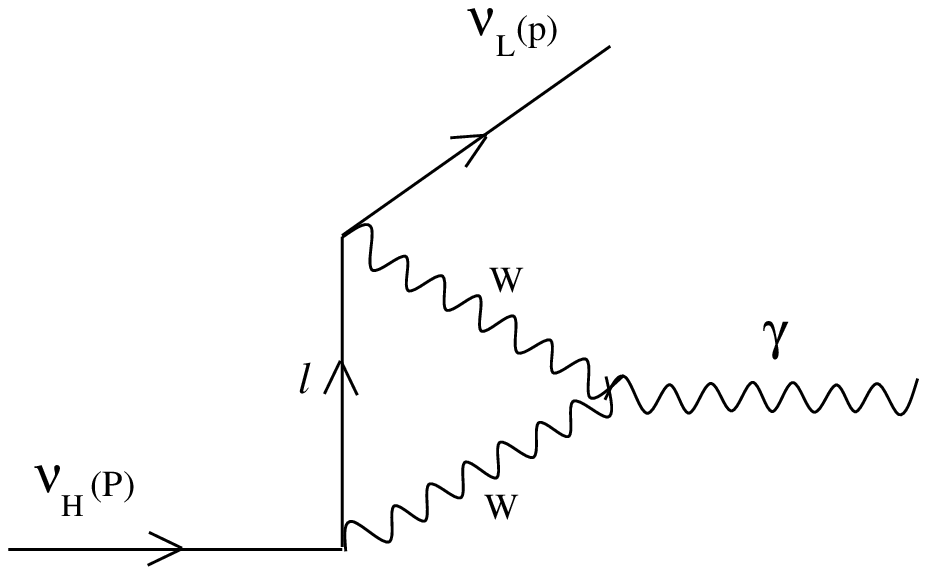,height=4.5truecm,width=5.5truecm}
\hskip 2cm
\epsfig{file=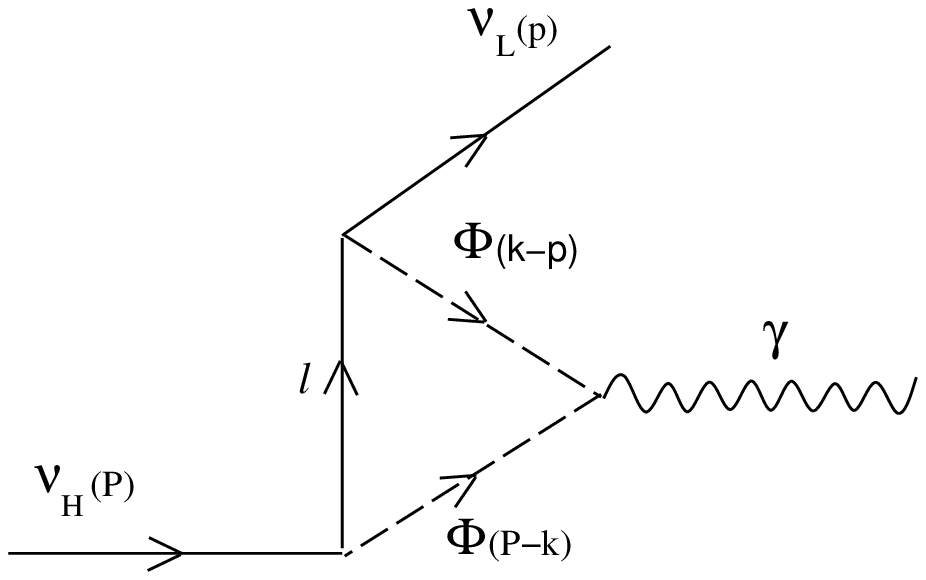,height=4.5truecm,width=5.5truecm}
\vskip 2mm
\hskip 2cm Fig.~1a \hskip 7cm Fig.~1b
\figskip
\epsfig{file=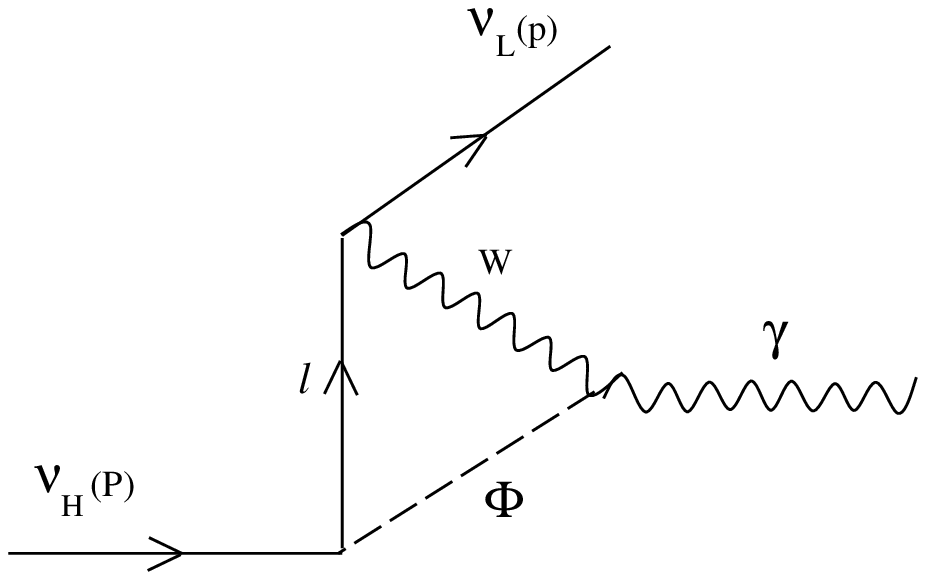,height=4.5truecm,width=5.5truecm}
\hskip 2cm
\epsfig{file=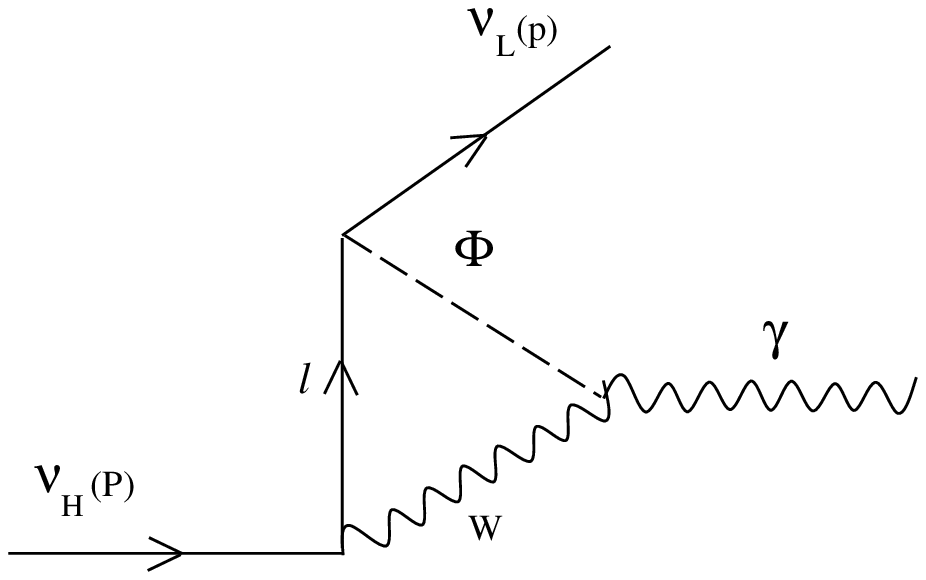,height=4.5truecm,width=5.5truecm}
\vskip 2mm
\hskip 2cm Fig.~1c \hskip 7cm Fig.~1d
\figskip
}
and two in Figs.~2a-b:

\vbox{
\figskip
\epsfig{file=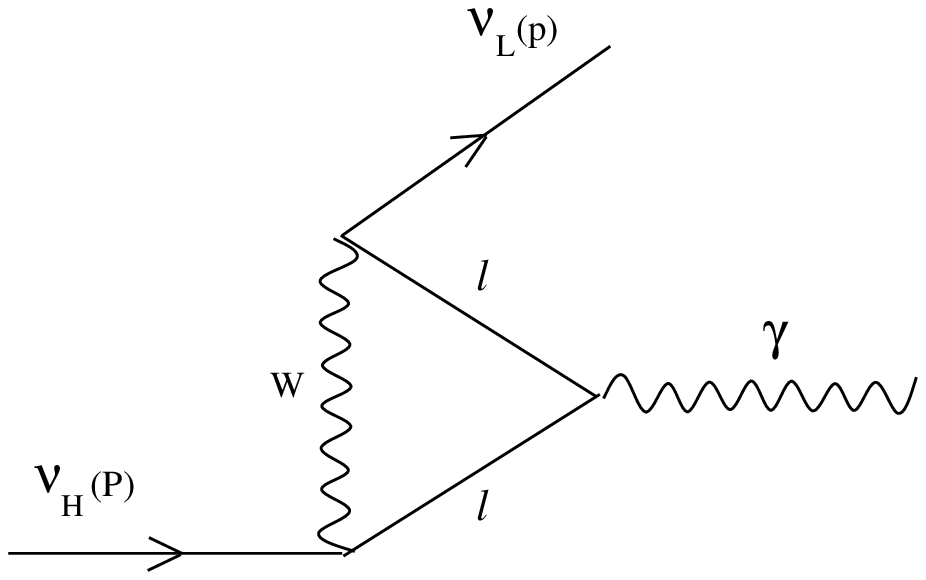,height=4.5truecm,width=5.5truecm}
\hskip 2cm
\epsfig{file=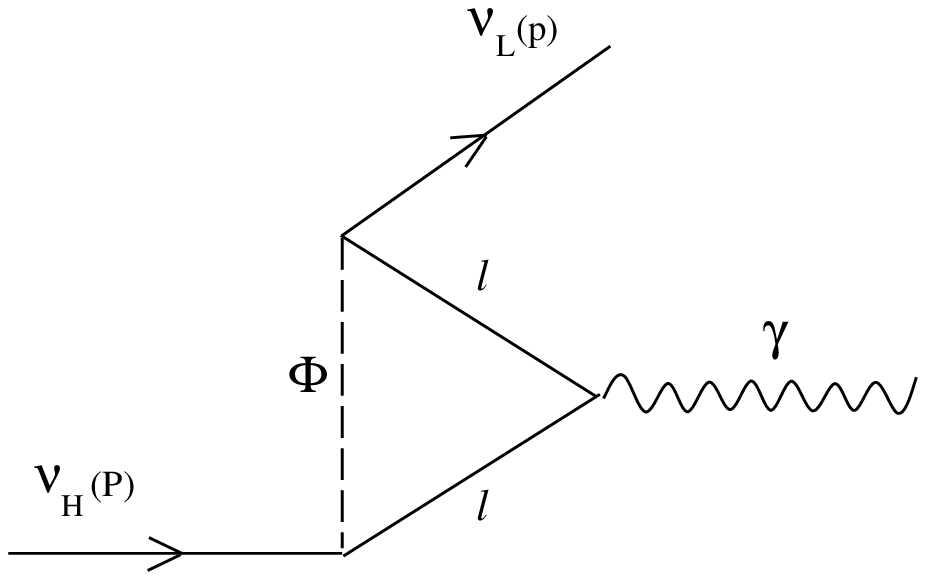,height=4.5truecm,width=5.5truecm}
\vskip 2mm
\hskip 2cm Fig.~2a \hskip 7cm Fig.~2b
\figskip
}
Each one is
gauge-dependent but it turns out that the $\xi$ dependence cancels out for
each group of diagrams separately, yielding the overall gauge independence
of the physical process.

We shall give the results in the 't Hooft-Feynman gauge $\xi = 1$.

For each diagram, the corresponding amplitude $\cal A$ is written in terms of
the effective  vertex $\Gamma_\mu$
\begin{equation}
i\,{\cal A}_{\nu_H \rar \nu_L\ \gamma} =
(-i\, e)\left(\frac{i g}{2 \sqrt{2}}\right)^2
\sum_\ell U_{L\ell} U^*_{H\ell}\;\overline{u}(p)\,\Gamma^\mu(\ell)\,u(P)
\;\varepsilon^*_\mu(q),
\end{equation}
where the $u$'s are the (Dirac) spinors of the two neutrinos,
$\varepsilon^*_\mu$ is the photon polarization, $e$ the charge of the
electron and $g$ the $SU(2)_L$ coupling constant. One has $G_F/\sqrt{2} =
g^2/8M_W^2$.

The ultraviolet  divergences are handled via the procedure of dimensional
regularization,  going to $n=4-\epsilon$ dimensions.

The mass $m$ of the lightest (outgoing) neutrino is always neglected, such
that the results depend on the mass $M$ of the incoming neutrino, the mass
$M_W$ of the $W$ gauge boson, and the masses $m_\ell$ of the internal
fermions, which will always appear in the dimensionless ratio
\begin{equation}
r_\ell = \frac{m_\ell^2}{M_W^2}.
\label{equation:rl}
\end{equation}
After expressing the amplitude for each diagram in terms of two-dimensional
parametric integrals, we restrict ourselves in this section to the case of
a real outgoing photon,
for which, due to $q^\mu \varepsilon^*_\mu =0$
and to the conservation of the electromagnetic current, only the magnetic
form factor proportional to $i\sigma_{\mu\nu}\,q^\nu$ in the effective
vertex contributes (see for example \cite{ChengLee}\cite{MohapatraPal}).
The integration over the Feynman parameters is made simpler by neglecting
$M^2/M_W^2$ in the denominators.

\subsection{Computation of the six diagrams}

\paragraph {$\bullet$ Diagram 1a}
The corresponding effective vertex $\Gamma^\mu_{1a}$ writes
\begin{equation}
\Gamma^\mu_{1a} = \frac{1}{8 \pi^2} (1+\gamma_5) \int_0^1 dx \int_0^{1-x} dy
\frac{{\cal N}_{1a}^\mu}{{\cal D}_1(\ell)}
\end{equation}
where
\begin{equation}
{\cal D}_1 (\ell) =M_W^2 [(1-x) +r_\ell x] -M^2 x y -q^2 \,y(1-x-y)
\end{equation}
and ($\gamma$ is the Euler constant $\gamma \approx .577$)
\bea
{\cal N}_{1a}^\mu &=& \Biggl\{[2(1-x)(1-y)+y] M^2-2 [(1-x)(1-y) +y^2]\,q^2
\Biggr .\cr
& &\Biggl .
+6{\cal D}_1(\ell)\left[\frac{2}{\epsilon}+\ln{4\pi}-\gamma -\frac{1}{2}
-\ln\frac{{\cal D}_1(\ell)}{\Lambda^2}\right]\Biggr\} \gamma^\mu\cr
& & + 2M\left\{y (1-2y) P^\mu +[2y^2 -(1-x)(1+2y)] p^\mu \right\}.
\label{eq:N1a}
\eea
In (\ref{eq:N1a}) and in the rest of the paper, $\Lambda$ is an arbitrary scale
coming from the dimensional regularization.

By translational invariance, $\Gamma^\mu$ depends only on the four-momentum
transfer $q^\mu$ and not on $P^\mu$; the latter may be projected onto the
basis formed by three independent four-vectors $i\, \sigma^{\mu\nu} q_\nu$,
$q^\mu$, and $\gamma^\mu$, using the following relation valid for $m=0$:
\begin{equation}
2\,\overline{u}(p)(1+\gamma_5)P^\mu u(P)=
\overline{u}(p) (1+\gamma_5)\left(i\,\sigma^{\mu \nu} q_\nu +
M \gamma^\mu + q^\mu\right)\,u(P).
\label{eq:proj}
\end{equation}
This yields

\vbox{
\bea
{\cal N}_{1a}^\mu &=&
i\, M\sigma^{\mu\nu} q_\nu [x-1 -y(1-2x)] +\qsl q^\mu [1-x+3y-2xy-4y^2]\cr
&+& \Biggl\{M^2[1-x-2y(1-2x)]-2q^2 [y^2+(1-x)(1-y)]\Biggr .\cr
&+& \Biggl . 6{\cal D}_1(\ell)\left[\frac{2}{\epsilon}+\ln{4\pi}-\gamma
-\frac{1}{2} -\ln\frac{{\cal D}_1(\ell)}{\Lambda^2}\right]\Biggr\} \gamma^\mu.
\label{eq:ND1a}
\eea
}
The $x,y$ integrations for the pure magnetic term yield the contribution of
diagram 1a to the decay amplitude $\nu_H\rar \nu_L \ \gamma$
\begin{equation}
{\cal A}_{1a}= {\cal A}_0 \sum_\ell U_{H\ell} U^*_{L\ell} F_{1a}(\ell),
\label{eq:A1b}
\end{equation}
where
\begin{equation}
{\cal A}_0 =\frac{G_F}{\sqrt{2}} \frac{e}{8\pi^2}
\overline{u}(p)\,M (1+\gamma_5)
 i \sigma^{\mu \nu} q_\nu \;u(P)\;\varepsilon^{*}_\mu(q).
\label{eq:A0}
\end{equation}
One gets
\begin{equation}
F_{1a}(\ell) =
\frac{r_\ell^2 (1-3r_\ell) \ln r_\ell}{2(r_\ell-1)^4}
+ r_\ell \left[\frac{7}{12 (r_\ell-1) } +\frac{2}{(r_\ell-1)^2}
+\frac{1}{(r_\ell-1)^3} \right] -{7\over 12}.
\label{eq:F1a}
\end{equation}
The singularities of $F_{1a}(\ell)$ at $r_\ell=1$ are fake:
$F_{1a}(\ell) = -5/12$ for $r_\ell =1$.
 
Formula (\ref{eq:F1a}) is in agreement with  similar calculations
\cite{ChengLee} for $\mu^- \rar  e^-\ \gamma$  at the limit $r_\ell \rar 0$,
where only the linear term in $r_\ell$ was kept and the logarithmic term
neglected.

If $m$ were not neglected, the $M(1+\gamma_5)$ term in (\ref{eq:A0})
would be simply replaced by $M(1+\gamma_5) +m(1-\gamma_5)$.
If we keep $M^2xy$ in ${\cal D}_1 (\ell)$, we will still obtain
explicit analytic forms for the $F$'s but the results will be complicated
and not illuminating.

\paragraph {$\bullet$ Diagram 1b}

Writing in a similar way
\begin{equation}
\Gamma_{1b}^\mu(\ell) = \frac{1}{8\pi^2}(1+\gamma_5)
\int_0^1 dx \int_0^{1-x}dy \frac{{\cal N}_b^\mu}{{\cal D}_1 (\ell)},
\end{equation}
one finds
\bea
{\cal N}_{1b}^\mu &=&
r_\ell\Biggl\{ M(1-y)\left[(2y-1)P^\mu +(1- 2x - 2y)p^\mu\right]\cr
& &+ {\cal D}_1(\ell)\,\left[\frac{2}{\epsilon}+ \ln{4\pi} -\gamma-\frac{1}{2}
- \ln\frac{{\cal D}_1(\ell)}{\Lambda^2} \right] \gamma^\mu \Biggr \}.
\eea
The use of (\ref{eq:proj}) transforms the above expression into
\bea
{\cal N}_{1b}^\mu &=&
r_\ell \Biggl\{
i\, M\sigma^{\mu\nu} q_\nu x(y-1)  + \qsl q^\mu (y-1)(1-x-2y)\cr
& &+ \left( M^2 x(y-1) + {\cal D}_1(\ell)\left[\frac{2}{\epsilon} +\ln{4\pi}
-\gamma - \frac{1}{2} - \ln\frac{{\cal D}_1(\ell)}{\Lambda^2} \right]\right)
\gamma^\mu
\Biggr \},
\label{eq:ND1b}
\eea
and, after performing the parametric integration of the purely magnetic
term one obtains
\begin{equation}
F_{1b}(\ell) =  \frac{r_\ell^2 (r_\ell-2) \ln r_\ell}{2(r_\ell- 1)^4}
+ r_\ell \left[-\,\frac{1}{3 (r_\ell-1)} - \,\frac{1}{4 (r_\ell-1)^2}
+\frac{1}{2(r_\ell-1)^3} \right].
\label{eq:F1b}
\end{equation}
The singularities of $F_{1b}(\ell)$ at $r_\ell=1$ are again only apparent;
in fact $F_{1b}(\ell) =-1/8$ for $r_\ell=1$.

The computations proceed along the same way for the other diagrams.
\paragraph{$\bullet$ Diagram 1c}
\begin{equation}
{\cal N}_{1c}^\mu =
i\, M\sigma^{\mu\nu} q_\nu (x+y -1) + \qsl q^\mu(1-x-y) +
\left( M^2(x-1) + m_\ell^2 \right) \gamma^\mu
\label{eq:ND1c}
\end{equation}
gives after the integrations over $x$ and $y$
\begin{equation}
F_{1c}(\ell) =
\frac{-r_\ell^2 \ln r_\ell}{2(r_\ell-1)^3} +
r_\ell \left[\frac{1}{4 (r_\ell-1)} +\frac{1}{2 (r_\ell-1)^2} \right]
- \frac{1}{4}.
\label{eq:F1c}
\end{equation}

\paragraph {$\bullet$ Diagram 1d}
\begin{equation}
{\cal N}_{1d}^\mu  = m_\ell^2\gamma^\mu
\label{eq:ND1d}
\end{equation}
yields
\begin{equation}
F_{1d}(\ell)  = 0.
\label{eq:F1d}
\end{equation}

\paragraph {$\bullet$ Diagram 2a}

Calling
\begin{equation}
{\cal D}_2 (\ell)=
M_W^2 x + m_\ell^2 (1- x) -M^2 x y -q^2 \,y(1-x-y)
\label{eq:D2}
\end{equation}
one has
\bea
{\cal N}_{2a}^\mu &=&
2 i\, M\sigma^{\mu\nu} q_\nu x(y-1) + 2\qsl q^\mu (1-y)(x+2y)\cr
& & +\Biggl\{ - 2m_\ell^2 + 2q^2 (y-1)(x+y)
- 2 {\cal D}_2 (\ell)\left[\frac{2}{\epsilon} + \ln{4\pi} -\gamma
-\frac{1}{2} - \ln \frac{{\cal D}_2 (\ell)}{\Lambda^2} \right]
\Biggr\} \gamma^\mu,
\cr
& &
\label{eq:ND2a}
\eea
and
\begin{equation}
F_{2a}(\ell) =
\frac{r_\ell (2r_\ell-1) \ln r_\ell}{(r_\ell-1)^4}
+ r_\ell \left[\frac{2}{3(r_\ell-1)} -\frac{3}{2 (r_\ell-1)^2}
-\frac{1}{(r_\ell-1)^3} \right]
- \frac{2}{3}.
\label{eq:F2a}
\end{equation}

\paragraph {$\bullet$ Diagram 2b}

\bea
{\cal N}_{2b}^\mu  &=&
r_\ell \Biggl\{
i\, M\sigma^{\mu\nu} q_\nu [x(1+y)-1] + \qsl q^\mu [1-x(1+y)- 2y^2]\cr
& &+ \left( -m_\ell^2 + M^2\,x + q^2 y(x+y-1)
-{\cal D}_2 (\ell)\left[ \frac{2}{\epsilon} + \ln{4\pi} -\gamma
- \frac{1}{2} -\ln \frac{{\cal D}_2 (\ell)}{\Lambda^2}\right]\right)
\gamma^\mu \Biggr\}
\cr
& &
\label{eq:ND2b}
\eea
yields
\begin{equation}
F_{2b}(\ell)  = 
\frac{r_\ell(2-r_\ell) \ln r_\ell}{2(r_\ell-1)^4}
+r_\ell \left[\frac{-5}{12 (r_\ell-1)} +\frac{3}{4 (r_\ell-1)^2}
-\frac{1}{2(r_\ell-1)^3} \right].
\label{eq:F2b}
\end{equation}

\subsection{Cancelation of the ultraviolet divergences}
\label{subsec:UV}

All terms that are $\ell$-independent do not
contribute to the amplitude because of the unitarity of ${\cal U}_{lep}$;
this is in particular the case of the (divergent) terms
$(2/\epsilon + \ln{4\pi} -\gamma -1/2)$ in the diagrams 1a and 2a.

The only two remaining divergent diagrams are 1b and 2b; however
the coefficients of their ($\ell$-dependent) divergent terms exactly cancel,
ensuring the finiteness of the final result.

\subsection{Result for the total amplitude of $\boldsymbol{\nu_H \rar \nu_L \
\gamma}$}

Dropping the constants $(-7/12), (-1/4), (-2/3$) in (\ref{eq:F1a}),
(\ref{eq:F1c}) and (\ref{eq:F2a}) which, being  $\ell$-independent,
do not contribute to the decay amplitude (see above), we obtain for
the sum of the six contributions
$\sum_{\ell}U_{H\ell} U^*_{L\ell} [ F_{1.a\cdots d}(\ell)
+ F_{2.a,b} (\ell)]$ the expression
\begin{equation}
{\cal A}_{\nu_{H}\rar \nu_{L} \ \gamma} = \frac{3}{4}{\cal A}_0
\sum_\ell U_{H\ell} U^*_{L\ell}\;\frac{r_\ell}{(1-r_\ell)^3}
\left[1-r^2_\ell +2r_\ell\ln r_\ell\right],
\label{eq:A}
\end{equation}
where ${\cal A}_0$ has been defined in (\ref{eq:A0}).
Our result (\ref{eq:A}) agrees with  formula (10.28) for the function $f(r)$
in reference \cite{MohapatraPal} (where the three irrelevant constants
mentioned above are kept).

The corresponding decay rate is
\begin{equation}
\Gamma_0\equiv \Gamma_{\nu_{H}\rar \nu_{L} \ \gamma}=
\frac{G_F^2 M^5}{192 \pi^3}
\left(\frac{27\alpha}{32 \pi}\right)
\left|\sum_\ell U_{H\ell} U^*_{L\ell} \;\frac{r_\ell}{(1-r_\ell)^3}
\left[1-r^2_\ell +2r_\ell\ln r_\ell\right]\right|^2.
\label{eq:Gamma0}
\end{equation}
With the assumptions about ${\cal U}_{lep}$ and the corresponding mixing
angles mentioned in the introduction, one finds for $M \approx
5\;10^{-2}\;eV$
\begin{equation}
\Gamma_{\nu_{H}\rar \nu_{L} \ \gamma} \approx 10^{-44}/\text{year}.
\end{equation}
This is to be compared with the experimental lower limit found in
\cite{Vanucci}.

The detectability of this decay and its relevance for astronomy has been
emphasized for example in \cite{RujulaGlashow}.

\section{The decay $\boldsymbol{\nu_H \rar \nu_L \ e^+ \ e^-}$}
\label{sec:ee}

If kinematically allowed, this decay is governed at tree level by the diagram
of Fig.~3, and at the one-loop level by ten diagrams: the six previously
considered in Figs.~1,2 where the photon, now off-mass-shell, decays into
an electron-positron pair, and the four box diagrams of Fig.~4 in which the
$W^+-W^-$ pair is converted into the $e^+-e^-$ pair.

\vbox{
\figskip
\centerline{\epsfig{file=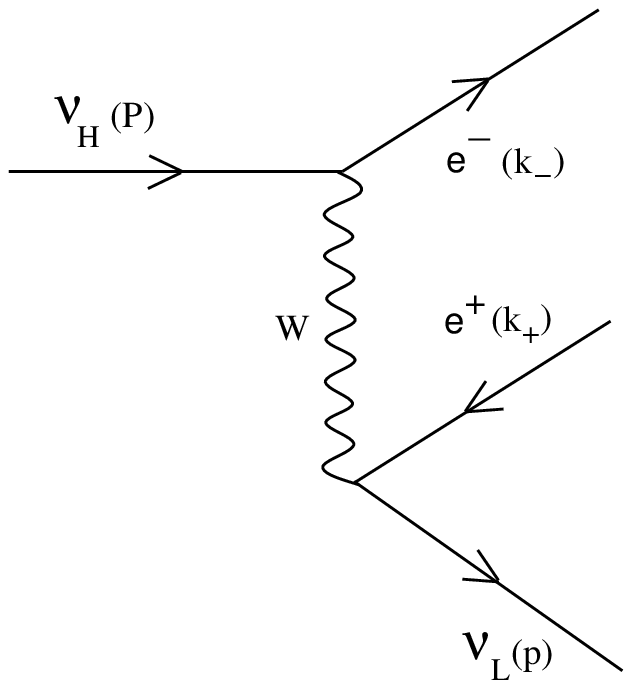,height=5truecm,width=5.5truecm}}
\vskip 2mm
\centerline{Fig.~3}
\figskip
}
The tree amplitude
\begin{equation}
{\cal A}_{tree} = \frac{G_F}{\sqrt{2}} U^*_{H e} U_{L e}\;
\overline{u}(k_{-}) \gamma^\mu (1-\gamma_5) u(P)\;
\overline{u}(p)\gamma_\mu (1-\gamma_5) v(k_{+})
\end{equation}
can be recast, by a Fierz transformation and using the unitarity of ${\cal
U}_{lep}$, into
\begin{equation}
{\cal A}_{tree} = (-1)^2 \frac{G_F}{\sqrt{2}}
\sum_{j= \mu, \tau} U^*_{H j} U_{L j}\;
\overline{u}(p) \gamma^\mu (1-\gamma_5) u(P)
\;\overline{u}(k_{-}) \gamma_\mu (1-\gamma_5) v(k_{+}).
\label{eq:Atree}
\end{equation}
As for the one-loop corrections, a careful examination of all the terms
in (\ref{eq:ND1a}), (\ref{eq:ND1b}), (\ref{eq:ND1c}), (\ref{eq:ND1d}),
(\ref{eq:ND2a}) and (\ref{eq:ND2b}) for the six vertices
$\Gamma^\mu_{1a-d}(\ell),\Gamma^\mu_{2a,b}(\ell)$ shows that the dominant
behavior comes from the  $q^2$ term  in (\ref{eq:ND2a}) corresponding to
Fig.~2a;  it exhibits a $\ln r_\ell \rar\infty$ for $r_\ell \to 0$
contribution, reflecting mass singularities (or infrared divergences) of the
loop integrals.

We can track down this divergent behavior by examining the integration
limits $x=0$ and $x=1$ of the denominators ${\cal D}_{1,2}(\ell)$.
When $r_\ell = 0$, an infrared-like divergence occurs if the numerators
${\cal N}_2^\mu(\ell)$ lack an $x$ term to cancel the $x=0$ integration limit
of the $xM_W^2$ term in the denominator ${\cal D}_2(\ell)$.
This happens with the $2y(y-1) q^2$ term of ${\cal N}_{2a}^\mu(\ell)$ in
(\ref{eq:ND2a}).

This infrared-like divergence, which arises when there are two massless
($r_\ell =0$) internal fermions in the loop, has been noticed a
long time ago in the computation of the neutrino charge radius \cite{BLBIM}.

Compared to $\ln\,r_\ell$, all other terms are negligible because they
are strongly damped by powers of $r_\ell$, or $r^n_\ell \, \ln\,r_\ell,
\text{where}\ n > 0$ and $r_\ell <10^{-3}$. Thus Fig.~2b is damped by
$r_\ell \ln\,r_\ell$, and the four diagrams \hbox{of Fig.~1} are all
strongly damped since an infrared-like divergence cannot occur here:
\hbox{the $x=1$} integration limit of the $(1-x) M_W^2$ in the
denominator ${\cal D}_1(\ell)$ is systematically canceled by the
$(1-x)$ coming from the integration over the $y$ variable.
Explicit $x,y$ integrations of all six vertices
$\Gamma^\mu_{1a-d}(\ell),\Gamma^\mu_{2a,b}(\ell)$ confirm these features.

Similar considerations show that
the box diagrams of Fig.~4 share the same power suppression $r^n_\ell \,
\ln\,r_\ell$ as the five other diagrams of Figs.~1a-d and Fig.~2b.
The origin of this $r_\ell$ power suppression in all one-loop diagrams
except Fig.~2a can be traced back to the fact that they involve two
$(W,\Phi)$ propagators; only Fig.~2a and Fig.~2b have one, but the
latter nevertheless gets an $r_\ell$ suppression from the $\Phi$-fermion
couplings.

\vbox{
\figskip
\centerline{\epsfig{file=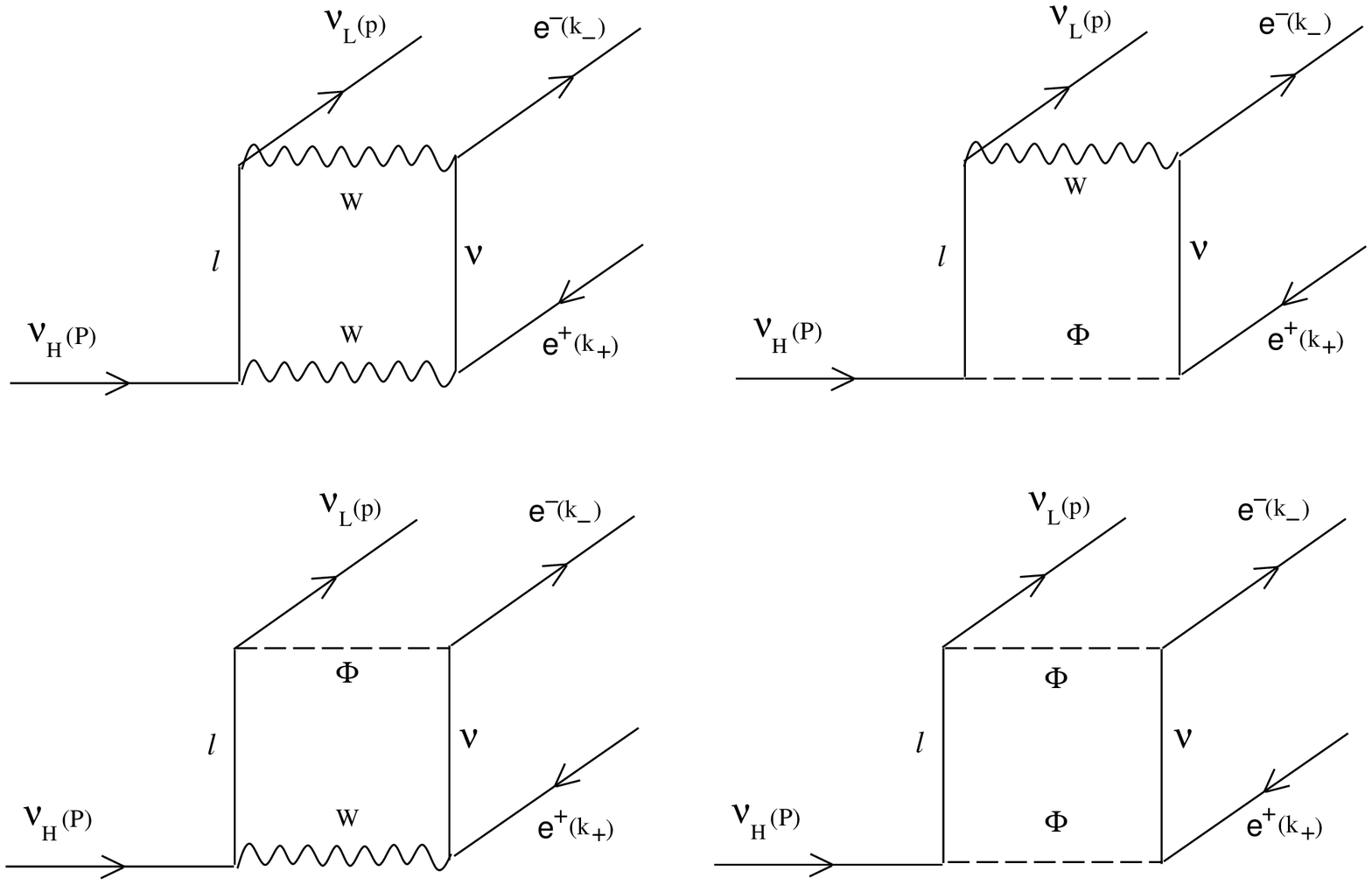,height=10truecm,width=16truecm}}
\vskip 2mm
\centerline{Fig.~4}
\figskip
}

To summarize, at one-loop, only the $2y(y-1)q^2$ term in (\ref{eq:ND2a})
yields an infrared-like divergence $\propto \ln r_\ell$ while all other terms
get damped by powers of $r_\ell$.

The leading $q^2\ln r_\ell$ term of Fig.~2a  in the $\nu_H-\nu_L-\gamma$
vertex cancels the photon propagator $1/q^2$ in Fig.~5 and yields an effective
local four-fermion coupling proportional to $G_F$. The leading contribution
to the one-loop radiative corrections  to the $\nu_H \rar \nu_L  \ e^+  \ e^-$
tree amplitude is accordingly found to be
\begin{equation}
{\cal A}_{rad} = \frac{G_F}{\sqrt{2}} \;{e^2\over 24 \pi^2}
\Bigl [\sum_{\ell} U^*_{H \ell} U_{L \ell}\;\ln r_\ell \Bigr]
\;\overline{u}(p) \gamma^\mu (1-\gamma_5) u(P)
\;\overline{u}(k_{-}) \gamma_\mu  v(k_{+}),
\end{equation}
which can be put, using again the unitarity of ${\cal U}_{lep}$
into a form similar to ${\cal A}_{tree}$ in (\ref{eq:Atree}):
\begin{equation}
{\cal A}_{rad} = \frac{G_F}{\sqrt{2}} \;{e^2\over 24 \pi^2}
\Bigl [\sum_{j=\mu,\tau} U^*_{H j} U_{L j}\;\ln \frac{m^2_j}{m_e^2}
\Bigr] \;\overline{u} (p) \gamma^\mu (1-\gamma_5) u(P)
\;\overline{u}(k_{-}) \gamma_\mu  v(k_{+}).
\end{equation}

\vbox{
\figskip
\centerline{\epsfig{file=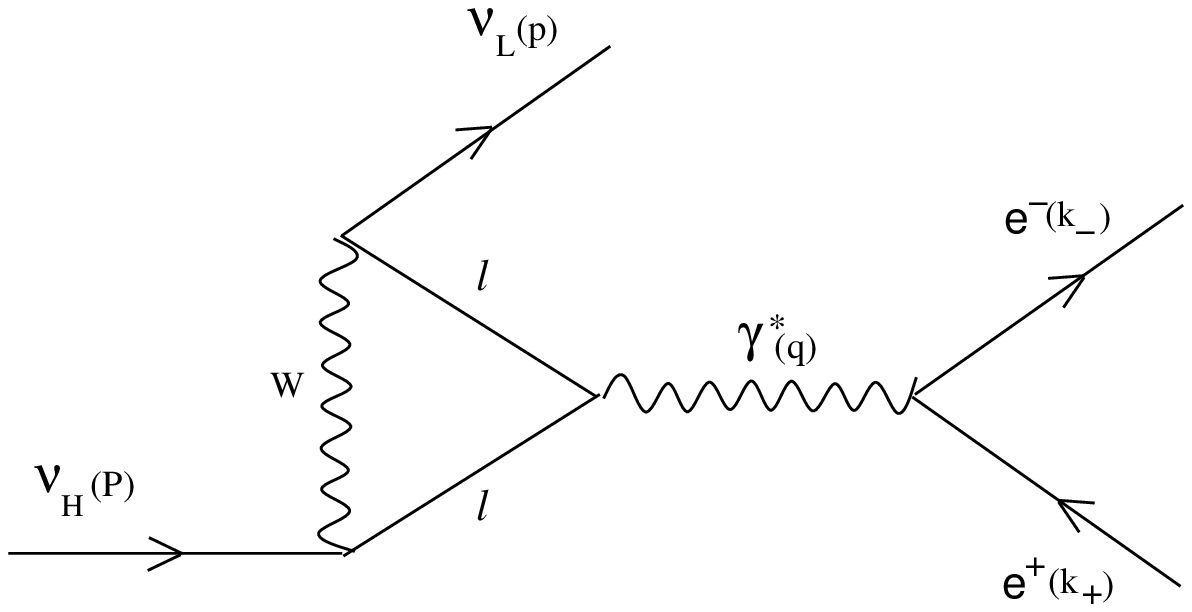,height=4.5truecm,width=7.5truecm}}
\vskip 2mm
\centerline{Fig.~5}
\figskip
}
The  sum ${\cal A}_{tree}+{\cal A}_{rad}\equiv {\cal B}$ is now easy to
manipulate when we consider the interference between ${\cal A}_{tree}$ and 
${\cal A}_{rad}$ in $|{\cal B}|^2 $ for the decay rate.
\begin{equation}
{\cal B} = \frac{G_F}{\sqrt{2}}
\;\overline{u} (p) \gamma^\mu (1-\gamma_5) u(P)
\;\overline{u}(k_{-}) \gamma_\mu (g_V-g_A\gamma_5)  v(k_{+}),
\end{equation}
with
\bea
g_V &=&  \sum_{j=\mu,\tau} U^*_{H j} U_{L j} \Bigl(1+ {\alpha\over 3\pi}
\ln \frac{m_j}{m_e}\Bigr),\cr
g_A &=&  \sum_{j=\mu,\tau} U^*_{H j} U_{L j}.
\label{eq:gvga}
\eea
{From} the amplitude ${\cal B}$, we compute \cite{HoKimPham} the decay rate
$\Gamma_1 \equiv \Gamma_{\nu_H \rar \nu_L  \ e^+  \ e^-}$  and find

\vbox{
\bea
\frac{d\Gamma_1}{d q^2}  &=& \frac{G_F^2}{192\pi^3} 
\frac{\sqrt {q^2(q^2-4m_e^2)}}{q^4 M^3} (M^2-q^2)^2\cr
& &\Bigl\{(g_V^2+g_A^2)\bigl[ q^2(M^2+2q^2) +2m_e^2 (M^2 -q^2) \bigr]
+6 m_e^2 q^2 (g_V^2-g_A^2)\Bigr\},
\eea
}
from which one gets
\begin{equation}
\Gamma_1 = \int_{4m_e^2}^{M^2} d q^2 \frac{d\Gamma_1}{d q^2 }
=  \frac{G_F^2 M^5}{192\pi^3}
\Bigl\{{g_V^2+g_A^2\over 2} G(x) + (g_V^2-g_A^2) H(x)\Bigr\},
\label{eq:Gamma1}
\end{equation}
where $x= m_e^2/M^2$, and $G(x), H(x)$ are the phase-space functions given by
\bea
G(x)&=& \left[ 1- 14 x - 2 x^2 -12 x^3\right] \sqrt{1-4x} + 24\,
x^2\,(1-x^2) \ln \frac{1+\sqrt{1-4x}}{1-\sqrt{1-4x}},\cr
H(x)&=& 2x(1-x)(1+6x)\sqrt{1-4x} + 12x^2 (2x-1-2x^2)
\ln \frac{1+\sqrt{1-4x}}{1-\sqrt{1-4x}}.
\label{eq:GH}
\eea
To this leading logarithmic radiative correction expressed by
$\approx \alpha \ln r$ in (\ref{eq:gvga},\ref{eq:Gamma1}), we may also add
the non-leading (simply $\alpha$, without $\ln r$) electromagnetic
correction to the $e^+\ e^-$ pair. This non-leading correction can
be obtained from the one-loop QCD correction to the well known $e^+\ e^-
\rar \text{quark-pair}$ cross-section, or the
$\tau \rar \nu_\tau +\text{quark-pair}$ decay rate found  in the literature
\cite{HoKimPham}; the only necessary change is the substitution
$\alpha_s \leftrightarrow 3\alpha/4$. Thus, in addition to $\Gamma_1$, we
have the non-leading contribution $\Gamma_2$
\begin{equation}
\Gamma_2 = \frac{G_F^2 M^5}{192\pi^3}
\left(\frac{3\alpha}{4\pi}\right) G(x) K(x,x);
\label{eq:Gamma2}
\end{equation}
the function $K(x,x)$ is tabulated in Table 14.1 of \cite{HoKimPham}.

We emphasize that $K(x,x)$ is a spectacular increasing function of $x$, 
acting in the opposite direction to the decreasing phase-space
function $G(x)$.

The present direct experimental limit on the mass of $\nu_\tau$ is \cite{PDG}
$m_{\nu_\tau} \leq 18.2$\,MeV; if we take, for example, the mass of the heavy
decaying neutrino to be $1.1$\;MeV, its lifetime is found to be
$\approx 10^{-2}$ year.

Other stronger limits (below $1\,MeV$) mainly come from cosmological arguments
\cite{DolgovDodelson}\cite{Kayser}.

Finally we note that the virtual weak neutral $Z^0$ boson replacing the virtual 
photon in Fig.~4 also contributes to $\nu_H \rar \nu_L \ e^+ \ e^-$.
However it can be safely discarded, being strongly damped by 
$q^2/M_Z^2$ due to the $Z^0$ propagator.

\section{Conclusions}
\label{section:conclusion}

The recent observation by the Super-Kamiokande collaboration of a clear
up--down $\nu_\mu$ asymmetry in atmospheric neutrinos is strongly
suggestive of $\nu_\mu \rar \nu_X$ oscillations, where $\nu_X$
may be identified with $\nu_\tau$ or even possibly a sterile neutrino.
These results have many important physical implications. In particular,
neutrino oscillations mean that neutrinos have a non-vanishing mass, which,
according to the new data, may be at least as heavy as $5\times 10^{-2}\,$
eV. If a neutrino $\nu_H$ has indeed a mass, it may not be stable
against decay and could in principle decay into a lighter neutrino,
$\nu_L$, through a cross-family electroweak coupling. We have
studied two such decay modes, $\nu_H \rar \nu_L\,\gamma$ and
$\nu_H \rar \nu_L\, e^+\, e^-$, and found that the latter, which,
in contrast to the former, arises at tree level and gets further enhanced
by large radiative one-loop corrections, is by far the dominant process
and may therefore be detectable provided that $\nu_H$ has a mass $>2\,m_e$.
A positive evidence for such decay modes would give a clear signal of the
onset of `new physics'.

\bigskip
\begin{em}
\underline {Acknowledgments}:
The work of Q.H.-K. was supported in part by the Natural Sciences and
Engineering Research Council of Canada.
B.M. and X.Y.P. thank J. Iliopoulos for bringing ref.~\cite{BLBIM} to their
attention.
\end{em}
\newpage\null
\listoffigures
\bigskip
\begin{em}
Figures 1--2 : One-loop diagrams for $\nu_H \rar \nu_L \ \gamma$;\l

Figure 3 : Tree diagram for  $\nu_H \rar \nu_L \ e^+ \ e^-$;\l

Figure 4: Box diagrams for $\nu_H \rar \nu_L \ e^+ \ e^-$;\l

Figure 5 : Leading radiative corrections to $\nu_H \rar \nu_L \ e^+ \  e^-$.
\end{em}
\newpage\null
\begin{em}

\end{em}


\begin{thebibliography}{50}
%
\bibitem{Superkamio}
       J. CONRAD \&  M. TAKITA; Talks at the XXIX International Conference
                   on High Energy Physics, Vancouver, Canada July 1998;\l
       Y. FUKUDA et al.: ``Evidence for Oscillation of Atmospheric
                          Neutrinos'', Phys. Rev. Lett. 81 (1998) 1562.

\bibitem{CKM}
       N. CABIBBO: ``Unitary symmetry and leptonic decays'',
                           Phys. Rev. Lett. 10 (1963) 531;\l
       M. KOBAYASHI \& T. MASKAWA:  ``$CP$-Violation in the Renormalizable
          Theory of Weak Interactions'', Prog. Theor. Phys. 49 (1973) 652.


\bibitem{Peccei}
       R.D. PECCEI: Summary talk at the XXIX International Conference
                    on High Energy Physics, Vancouver (Canada) July 1998.
\bibitem{Bugey}
       C. HAGNER {\em et al.}: ``Experimental search for the neutrino decay
         $\nu_3 \rightarrow \nu_j + e^+ + e^-$ and limits on neutrino
          mixing'', Phys. Rev. D52 (1995) 1343.

\bibitem{Petcov}
       S.T. PETCOV: ``The processes $\mu \rightarrow e +\gamma,\ \mu
                \rightarrow e+e+\bar e,\ \nu' \rightarrow \nu + \gamma$ in
                the Weinberg-Salam model with neutrino mixing'',
               Yad. Fiz. 25 (1977) 641 (Sov.J. Nucl. Phys. 25 (March 1977) 340);
               {\em Erratum}: Sov.J. Nucl. Phys. 25 (June 1977) 698.

\bibitem{PalWolfenstein}
       P.B. PAL \& L. WOLFENSTEIN: ``Radiative decays of massive neutrinos'',
             Phys. Rev. D 25 (1982) 766.

\bibitem{ChengLee}
       T.P. CHENG \& L.F. LI:  ``Gauge Theory of Elementary Particle Physics'',
            p. 426, Oxford University Press (New York) 1984.

\bibitem{MohapatraPal}
       R.N. MOHAPATRA \& P.B. PAL: ``Massive Neutrinos in Physics and
             Astrophysics'', p. 183,  World Scientific (Singapore) 1991.

\bibitem{Vanucci}
       C. BIRNBAUM {\em et al.}: ``An experimental limit on radiative decays
                   of solar neutrinos'', Phys. Lett. B 397 (1997) 143.

\bibitem{RujulaGlashow}
       A. de R\'UJULA \& S.L. GLASHOW: ``Galactic Neutrinos and uv
                        Astronomy'', Phys. Rev. Lett. 45 (1980) 942.

\bibitem{BLBIM}
       J. BERNSTEIN \& T.D. LEE: ``Electromagnetic form factor of the
                       neutrinos'', Phys. Rev. Lett. 11 (1963) 512;\\
       C. BOUCHIAT, J. ILIOPOULOS \& Ph. MEYER: ``Elastic neutrino
           scattering in renormalizable theories of weak and electromagnetic
           interactions without neutral currents'', Phys. Lett. B42 (1972) 91.

\bibitem{HoKimPham}
       Q. HO-KIM \& X.Y. PHAM: ``Elementary Particles and their
           Interactions. Concepts and Phenomena'',
                Springer (Berlin, Heidelberg) 1998.

\bibitem{PDG}
       ``Review of Particle Properties''; The European Physical Journal C 3
             (1998).

\bibitem{DolgovDodelson}
        A.D. DOLGOV \& I.Z. ROTHSTEIN: ``New Upper Limits on the
             Tau-Neutrino Mass form Primordial Helium Considerations'',
             Phys. Rev. Lett.  71 (1993) 476;\l
        S. DODELSON, G. GYUK \& M.S. TURNER: ``Is a Massive Tau Neutrino
                         Just What Cold Dark Matter Needs?'',
                    Phys. Rev. Lett. 72 (1994) 3754;\l
        A.D. DOLGOV, S.H. HANSEN \& D.V. SEMIKOZ: ``Impact of massive
             tau-neutrinos on primordial nucleosynthesis. Exact calculations'',
             Nucl.  Phys. B 524 (1998) 621;\l
        A.D. DOLGOV, S.H. HANSEN, S. PASTOR \& D.V. SEMIKOZ: ``Unstable
            massive tau-neutrino and primordial nucleosynthesis'',
           hep-ph/9809598 (revised Feb. 1999).

\bibitem{Kayser}
        B. KAYSER: ``Neutrino Mass and Oscillation'', hep-ph/9906244 (June
                   1999).
%
\end{thebibliography}
\end{document}